\documentclass[12pt]{article}
\usepackage{graphics}
\usepackage{amsmath}
\usepackage{amssymb}
\usepackage{graphicx}
\usepackage{makeidx}
\usepackage{mathrsfs}
\usepackage{amsfonts}
\usepackage[mathscr]{eucal}
\usepackage{amsmath}

\usepackage{graphics}

\DeclareGraphicsExtensions{.eps,.pdf,.jpg,.png}

\def\e{{\rm e}}
\def\e{\hbox{e}}
\def\d{\hbox {d}}

\def\ds{\displaystyle}
\def\RR{\vbox {\hbox to 8.9pt {I\hskip-2.1pt R\hfil}}}

\def\CC{{\rm C\hskip-4.8pt \vrule height 6pt width 12000sp\hskip 5pt}}
\def\cen{\centerline}
\def \rec#1{{1\over{#1}}}
\def\pni{\par\noindent}
\def\vsh{\smallskip}

\def\vsp{\vsh\pni} %% ie. \smallskip + \par

\DeclareGraphicsExtensions{.eps,.pdf,.jpg,.png}

\begin{document}

\cen{{\bf FRACALMO PRE-PRINT: \ http://www.fracalmo.org}}
% \vsh
\cen{{\bf The European Physical Journal, Special Topics, Vol. 193  (2011) 161--171}}
% \vsh
\cen{{\bf Special issue:   Perspectives on Fractional Dynamics and Control}}
%% \vsh
\cen{{\bf Guest Editors: Changpin LI  and Francesco MAINARDI }}
\vsh
\hrule
% \end{center}
%%%%%%%%%%%%%%%%%%%%%%%%%%%%%%%%%%%%%%%%%%%%%%%%%%%%%%%%%%%%%%%%%%%%%%%%
\vskip 0.50truecm
\font\title=cmbx12 scaled\magstep2
\font\bfs=cmbx12 scaled\magstep1
\font\little=cmr10
\begin{center}
{\title Models based on Mittag-Leffler functions}
 \\ [0.25truecm]
 {\title for anomalous relaxation in dielectrics}
 \\ [0.25truecm]
Edmundo CAPELAS DE OLIVEIRA$^{(1)}$ 
Francesco MAINARDI$^{(2)}$ and 
Jayme VAZ Jr.$^{(1)}$
\\[0.25truecm]
$\null^{(1)}$ {\little Department of Applied Mathematics, IMECC, University of Campinas,}
\\
 {\little 13083-869 Campinas, SP, Brazil} 
\\ 
{\little E-mail: capelas@ime.unicamp.br, vaz@ime.unicamp.br}
\\[0.25truecm]
$\null^{(2)}$ {\little Department of Physics, University of Bologna, and INFN}
\\ 
{\little Via Irnerio 46, I-40126 Bologna, Italy}
\\{\little Corresponding Author.   E-mail: francesco.mainardi@bo.infn.it}
\vskip 0.5truecm
{\bf Second Revised Version: February 2014}
\end{center}
%%%%%%%%%%%%5
\begin{abstract}
%% Insert your abstract here.
\noindent 
We revisit the Mittag-Leffler functions of a  real variable $t$, with one, 
two and three order-parameters $\{\alpha, \beta, \gamma\}$, 
as far as their Laplace transform pairs and complete monotonicity 
properties are concerned. 
These functions, subjected to the requirement to be completely monotone for $t>0$, are shown to be 
suitable models for non--Debye relaxation phenomena in dielectrics 
including as particular cases the classical models referred 
to as Cole--Cole, Davidson--Cole and Havriliak--Negami. 
We show 3D plots of the response functions and of the corresponding spectral distributions, 
keeping fixed one of the three order-parameters. 
\end{abstract}
%
% \maketitle

\vsp
{\it 2010 Mathematics Subject Classification (MSC)}:
26A33, 33E12,  44A10.
%% 26A33,  %%%%  (main);    Fractional derivatives and integrals
%% 33E12, %% Mittag-Leffler type functions
%% 44A10,  %% Laplace Transforms
\vsp
{\it 2010 Physics and Astronomy Classification Scheme (PACS)}:
%% 77.84.s %% Dielectric materials, 77.84.-s
77.22.Gm. %%% Dielectric relaxation, 77.22.Gm
\vsp
\newpage
\section*{Foreword to the second revised version, \\ February 2014}
The colleagues
Rudolf Gorenflo, Tibor K. Pogany   and \v{Z}ivorad Tomovski,
 when collaborating on use of the Prabhakar function, found a mistake in our use of the theorem by Gripenberg et al.   in Section 2.3 of the first version.
%% for stating the conditions of complete monotonocity. 
In this revised version we will properly apply the conditions of this theorem in order to improve our previous results. We  take this occasion to correct a number of misprints 
and improve Section 2.3. 
Furthermore, 
 we will  better rearrange  the original text   and update the bibliography.
The authors are thus very grateful to these colleagues for having pointed out
the deficiencies of the previous analysis whose final results, however, remain still valid but less general, as it will be shown in the following.

\section*{Foreword to the first revised version, \\ June 2011}
This E-print reproduces the revised version of the paper published in EPJ-ST, Vol. 193 (2011), pp. 161--171. 
The revision concerns the proper use of the terms {\it relaxation function}
 and {\it  response function} in the literature on dielectrics. 
 In the published paper, starting from Eq.(1.1), the authors had referred the inverse Laplace transform of the  
 complex  susceptibility as the relaxation function. This is not correct because the inversion provides the so-called response function as pointed out to
one of the authors (FM) by Prof. Karina Weron (KW) to whom the authors are 
very grateful. 
As a matter of fact the relationship between the response function and the relaxation function 
can better be clarified by their probabilistic interpretation investigated in several papers by KW.
As a consequence,  interpreting  the {\it relaxation function} as a {\it survival probability} $\Psi(t)$,
the {\it response function} turns out to be  the {\it probability density function} corresponding 
to the {\it cumulative probability function} $\Phi(t)= 1 - \Psi(t)$.
Then, denoting by $\phi(t)$ the response function, we have  
$$\phi(t) = - \frac{d}{dt}\Psi(t) =  \frac{d}{dt}\Phi(t)\,, \quad t\ge 0\,.$$ 
As KW has pointed out,  both functions have very different properties and describe different physical
magnitudes; only in the Debye (pure exponential) case the properties coincide. 
The relaxation function describes the decay of polarization whereas the response function its decay rate 
(the depolarization current).  However, for physical realizability, both functions are required 
to be  completely monotone with a proper spectral distribution so our analysis can be properly transferred 
from response functions to the corresponding relaxation functions, whereas the corresponding 
cumulative probability functions turn out to be Bernstein (or creep) functions,
that is  positive functions with a completely monotone derivative.    
\vsp   
In  the following  we denote the response function $\phi(t)$ with $\xi(t)$ in order to be consistent with our  notation
 $\widetilde \xi(s)$ with $s=i\omega$  
 for the   complex  susceptibility as a function of the frequency $\omega$.

\section{Introduction}
\label{intro}
It is well recognized that relaxation phenomena in dielectrics deviate more or less 
strongly from the classical Debye law for which the Laplace transform pair for complex 
 susceptibility ($s=-i\omega$) and response function ($t\ge 0$) reads in an obvious 
notation,
$$ 
\widetilde \xi_{\hbox{D}} (s)  = \frac {1}{1 + s}\,
\div \, \xi_{\hbox{D}} (t) =  \e ^{\,\ds -t}\,.
 \eqno(1.1)
$$
%%%%%%%
Here, for the sake of simplicity, we have assumed the frequency $\omega$ and the time $t$
normalized with respect to a characteristic frequency $\omega_D$ and  a corresponding 
relaxation time $\tau_D =1/\omega_D$.
\vsp
In the literature a number of laws  have been proposed to describe  the non-Debye (or  anomalous) 
relaxation phenomena in dielectrics, 
of which the most relevant ones  are referred to   Cole -- Cole (C-C), Davidson -- Cole (D-C) and
  Havriliak -- Negami (H-N) laws, see e.g., the classical books by Jonscher 
  \cite{Jonscher BOOK83,Jonscher BOOK96}.
  Several authors have investigated these laws from different points of view, including 
  Karina Weron and her associates, see e.g.
  \cite{Jurlewicz-Weron 2000,%
  Jurlewicz-Weron 2002,% 
  Jurlewicz-Weron-Teuerle 2008,%
  Stanislavsky-Weron EPL10,%
  Szabat-Weron-Hetman 2007,%
 Weron-et-al 2005,%  
  Weron-Kotulski 1996},
  %% Metzler and Klafter \cite{Metzler-Klafter 2004},
   Hilfer \cite{Hilfer 2002a,Hilfer 2002b} and  Hanyga and Seredy\'nska \cite{Hanyga JSP08}.   
% \vsp
In particular,  Hilfer  surveyed the analytical expressions in the frequency and time  domain 
for the main non-Debye relaxation processes and  provided 
the response functions corresponding to the complex frequency-dependent Cole-Cole, 
Davidson-Cole and Havriliak-Negami  susceptibilities in terms of Fox $H-$functions.
This class of  functions is quite general so  it includes the Mittag-Leffler functions 
that we prefer to use to  characterize the above laws in a more accessible way.  
%%%%%%%%%%%%%
\newpage
%%%%%%%%
\vsp
On the other hand, for linear systems, the connection between weak dissipativity and positive definiteness 
of the response functions as well as between monotone energy decay and complete 
monotonicity of the  response functions  
were discussed by Hanyga and Seredy{\'n}ska \cite{Hanyga JSP08}, and in references therein, 
in terms of functions of Mittag-Leffler type.
A subordination model of anomalous diffusion leading to 
the two-power-law relaxation responses  has been  proposed by Stanislavsky et al. \cite{Stanislavsky-Weron EPL10}, 
where the authors have presented a novel two-power relaxation law and shown its 
relationship to the H-N law by using functions of Mittag-Leffler type.
%\vsp
Moreover,
Nigmatullin and Ryabov \cite{Nigmatullin-Ryabov 1997},
 Novikov et al. \cite{Novikov MS05} and
 Sibatov  et al. \cite{Sibatov-et-al 2012}
  have discussed anomalous relaxation in dielectrics
providing   evolution equations with fractional derivatives to describe the 
relaxation of the C-C, C-D and H-N types in dielectrics.
%% The corresponding solutions  were presented in terms of Mittag-Leffler functions.
For a treatise on fractional relaxation in dielectrics we refer the reader to the recent book  by Uchaikin and Sibatov \cite{Uchaikin-Sibatov_BOOK2012}. 
%% The H-N response function has been pointed out in detail by Sibatov et al 
% \cite{Sibatov-Uchaikin E-print10}  who have  obtained an explicit expression and %%also  presented a comparison of numerical results with experimental data. 
\vsp
The purpose of this paper is to present a general model for anomalous relaxation in 
dielectrics that includes as particular cases the classical C-C, D-C and N-H laws.
Our model is still based on the Mittag-Leffler functions but depending on three order-parameters 
that, in view of their complete monotonicity in the time domain, ensure the existence of
a suitable spectrum of relaxation times required for the  physical  realizability.
On the realizability requirements the reader can be addressed, in addition to  \cite{Hanyga JSP08},
  to the treatise by Zemanian \cite{Zemanian BOOK72}.
 %\vsp
% In a subsequent paper we will investigate how to provide  the differential equation % of fractional order  that is expected to govern our general model, so generalizing the % approach and the results by  Novikov et al.\cite{Novikov MS05}   
%% and Uchaikin et al. et al. \cite{Uchaikin-et-al 2013}.
\vsp
The present paper is organized as follows. 
% \vsp
In section 2 we recall the definitions of the most common functions of Mittag-Leffler type, namely  
those depending  on  one, two and three positive order-parameters 
 $\{\alpha, \beta, \gamma\}$. 
For these functions we exhibit the corresponding Laplace transform pairs 
and we discuss their properties of complete monotonicity. 
We show that the complete monotonicity is ensured if 
 the independent variable is real and negative and the three positive order-parameters are subjected to the condition $\alpha\,\gamma \le \beta \le 1$ with $\alpha \le 1$,
 For other properties  on  the Mittag-Leffler functions we refer the reader mainly to  texts on Special Functions
and Fractional Calculus, including  {\it e.g.} 
\cite{Diethelm BOOK10,Hilfer BOOK02,Kilbas-Saigo BOOK04,%% 
Kilbas-et-al BOOK06,Kiryakova BOOK94,Mainardi BOOK10,%%
Mathai-Haubold BOOK08,Mathai-Saxena-Haubold BOOK10,Podlubny BOOK99},
 and to the recent survey by Haubold et al.
  \cite{Haubold-Mathai-Saxena_JAM2011}.
%% Podlubny (1999), Hilfer (2000), Kilbas, Srivastava \& Trujillo (2006), 
%% Mathai, Saxena \& Haubold (2008),   Mainardi (2010), Diethelm (2010) 
%% just to cite some of the most recent ones. 
In fact, as pointed out by Gorenflo and 
Mainardi \cite{Gorenflo-Mainardi CISM97,Mainardi-Gorenflo FCAA07}, 
functions of Mittag-Leffler type enter as solutions of many problems dealt  
with fractional calculus so that  they like to refer to the 
Mittag-Leffler function to as {\it the Queen function of Fractional Calculus}, in contrast with 
its role of a {\it Cinderella function}  played in the past.  
\vsp
In section 3 we show that  for special cases of the triplet 
$\{\alpha,\beta, \gamma  \}$ with $\alpha\gamma -\beta =0$
and $0\! <\! \alpha, \beta, \gamma \!\le \! 1$
our functions provide the response functions for the classical models of Cole--Cole
($\{0\! < \! \alpha \! = \! \beta \! <\! 1 , \gamma \!=\!1  \}$),
 Davidson--Cole
 ($\{\alpha =1, 0<\beta= \gamma <1  \}$),
   and Havriliak--Negami
 ($\{0<\alpha ,\! \gamma<1, \, \beta= \alpha \gamma \}$).
 %\vsp
 As a consequence,  we expect that in the more general case 
 $\alpha\, \gamma -\beta \le 0$   the corresponding Mittag--Leffler functions, being completely monotone, can provide 
  further models for processes of anomalous (non-Debye) relaxation in dielectrics.
For some study-cases, taking fixed two of the three order-parameters,
 we provide   3D plots of the responses functions 
 and of the corresponding spectral distributions, 
  in order to better visualize  the positivity and the variability  of the considered functions.  
\vsp
Finally, section 4   is devoted to conclusions and final remarks.
%%	together with a discussion on the direction of future work.
	
\section{The Mittag-Leffler functions}
\label{sec:1}

\subsection{Definitions}
\label{sec:1.1}
We start to recall  the definition in the complex plane
of the generalized Mittag-Leffler function introduced by Prabhakar \cite{Prabhakar 71},
known as Prabhakar function or 3-parameter Mittag-Leffler function 
$$ E_{\alpha,\beta}^{\gamma}(z) :=
\sum_{n=0}^{\infty} \frac{(\gamma)_n}{n!\Gamma(\alpha n+ \beta)}\, z^n\,, \;
\alpha, \beta, \gamma \in \CC \; \hbox{with} 
\; {\mbox{Re}} \{\alpha\} > 0, \eqno(2.1)$$
where
$$  (\gamma)_n = \gamma (\gamma+1)\dots (\gamma + n -1)= 
\frac{\Gamma(\gamma+n)}{\Gamma (\gamma)}\,.$$
It turns out to be an entire function of order $\rho = 1/{\mbox{Re}} \{\alpha\}$.
%%%
For $\gamma=1$ we recover the 2-parameter Mittag-Leffler function 
%(also known as Wiman or  Agarwal generalized Mittag-Leffler function) %%% $E_{\alpha,\beta}(z)$
$$ E_{\alpha,\beta}(z) :=
\sum_{n=0}^{\infty} \frac{z^n}{\Gamma(\alpha n+ \beta)}\,,  \eqno(2.2)$$
and for $\gamma=\beta=1$ we recover the standard Mittag-Leffler function
 $$ E_{\alpha}(z) :=
\sum_{n=0}^{\infty} \frac{z^n}{\Gamma(\alpha n+ 1)}\,. \eqno(2.3)$$
  Henceforth we consider all three parameters to be real with $\alpha >0$. 
%%%  
  \newpage
  %%%%
\subsection{The Laplace transform pairs}
\label{sect:1.2}
Let us now consider  the relevant formulas of Laplace transform pairs related to the above three functions already known in the literature when
the independent variable is  real of type $at$ where $t>0$ may be  interpreted as time and $a$ as  a certain constant of frequency dimensions.
For the sake of convenience we adopt the notation $\div$   
to denote the juxtaposition of a function of time $f(t)$ with
its Laplace transform 
$\widetilde f(s) =  \int_0^\infty \! \e^{-st} f(t) \, dt$.
We have 
$$
t^{\beta-1}\,{\ds E_{\alpha,\beta}^{\gamma}\left( at^\alpha\right)} 
\,\div \,
 {\ds \frac{s^{-\beta}}{\left( 1-a s^{-\alpha}\right)^\gamma}}
 = {\ds \frac{s^{\alpha \gamma -\beta}}{\left( s^{\alpha} -a\right)^\gamma}}
\,. 
\eqno(2.4)$$
$$
t^{\beta-1}\,{\ds E_{\alpha,\beta}(at^\alpha)} \,\div \,
{\ds \frac{s^{-\beta}}{1-  as^{-\alpha}}}
={\ds \frac{s^{\alpha -\beta}}{  s^{\alpha}-a }}
\,, 
\eqno(2.5)$$
$$
{\ds E_{\alpha}(at^\alpha)} \,\div \,
 {\ds \frac{s^{ -1}}{ 1- a s^{-\alpha }}}
={\ds \frac{s^{\alpha -1}}{ s^{\alpha}-a } }
\,. 
\eqno(2.6)$$
  All the three parameters $\alpha, \beta, \gamma$ are required to be positive
  with $\mbox{Re} \{s\} >0$ and $\left | a s^{-\alpha}\right | <1$, see {\it e.g.}
  \cite{Kilbas-et-al BOOK06,Mainardi BOOK10}.
\vsp
Let us report the proof  of the general Laplace transform pair Eq.(2.4). 
Substituting the series representation of the Prabhakar
 function in the Laplace transformation yields the identity 
$$
\int_0^\infty \e^{\ds -st}\, t^{ \beta-1}\,{\ds E_{\alpha,\beta}^{\gamma}(at^\alpha)\, {\mbox{d}}t} =
s^{-\beta} \,
\sum_{n=0}^{\infty} 
\frac{\Gamma(\gamma+n)}{\Gamma (\gamma) n!}\left(\frac{a}{s^\alpha} \right)^n
 \,.  
\eqno(2.7)
$$
On the other hand,  recalling the binomial series with $\gamma>0$, we have 
$$
(1+z)^{-\gamma} =
\sum_{n=0}^\infty \frac{\Gamma(1-\gamma)}{\Gamma(1-\gamma -n) n!}\, z^n =
\sum_{n=0}^\infty (-1)^n\,\frac{\Gamma(\gamma +n)}{\Gamma(\gamma) n!}\, z^n\,. 
\eqno(2.8)
$$
Comparison of Eq.(2.7) with Eq.(2.8) yields the Laplace transform pair 
in Eq.(2.4) and consequently the pairs in Eqs.(2.5)-(2.6).
We recognize that the condition $\beta >0$ is now necessary to ensure that the function be absolutely integrable close to the origin (hence locally integrable in $\RR^+$)  so the corresponding Laplace transform goes to zero as $s^{-\beta}$
for $\mbox{Re} \{s\} \to +\infty$.
%% Of course we also recognize the necessity of  $\mbox{Re} \{s\} >0$ with 
%% $\left | a s^{-\alpha}\right | <1$  and $\alpha, \gamma >0$.
In conclusion for the Laplace transform provided in  Eqs.(2.4) all the three parameters
$\alpha, \beta, \gamma$  are required to be positive.
\vsp
We also note that   in general with five positive parameters $\alpha, \beta, \gamma, \delta, \rho$ 
 the function $ {\ds t^{\rho -1}\, E_{\alpha, \beta}^{\gamma} \left(at^\delta\right)}$ 
has a  Laplace transform expressed in terms of a transcendental function of 
 Wright  hypergeometric type, see Eq. (11.10) in 
\cite{Haubold-Mathai-Saxena_JAM2011}.
Only for the special case $\rho =\beta$ and $\delta=\alpha$ the Wright function
turns to be expressed in terms of the algebraic function in the RHS of Eq.(2.4).
%%%%%%%%%
\subsection{Complete monotonicity}
Let us recall that a real non-negative function $f(t)$ defined  for $t \in {\mathbb{R}}^+$ 
is said to be completely monotone
(CM)  if it possesses derivatives $f^{(n)}(t)$ for all $n = 0,1,2,3,. .$ 
that  are alternating in sign, namely 
 $$ (-1)^n f^{(n)}(t) \ge 0 \, \quad \ t > 0\,.\eqno(2.9)$$
The limit $f^{(n)} (0^+)  = {\displaystyle \lim_{t\to 0^+}} f^{( n)}(t)$ 
finite or infinite exists. 
For the existence of the Laplace transform of $f(t)$ we require
that the function be locally integrable in $\RR^+$.   
Thus, for the Bernstein theorem that states a necessary and sufficient condition for the CM,   the function  can be expressed  as a real Laplace transform 
of non-negative (generalized) function, namely
$$ f(t) = \int_0^\infty \e^{\ds -rt}\,  K(r)\, \d r\,,\;  K(r) \ge 0\,, \; t\ge 0 \,.\eqno(2.10) $$
For more details see {\it e.g.} the survey by Miller \& Samko \cite{Miller-Samko 01}.
By the way, the above representation with the determination of  such non-negative 
function $K(r)$  is a standard method to prove the CM of a given function defined 
in the positive real semi-axis  $\RR^+$. In physical applications the function $K(r)$ 
is usually referred to as  the {\it spectral distribution}, 
in that it is related to the fact that the process governed by the function 
$f(t)$ with $t\ge 0$
can be expressed in terms of a continuous distribution of elementary (exponential) relaxation processes with  frequencies $r$ on the whole range $(0,\infty)$.
In the case of the pure exponential $f(t)= \exp(-\lambda t)$ with a given relaxation frequency $\lambda >0$ we have 
$K(r;\lambda)= \delta(r-\lambda)$. 
 \vsp
  Since $\widetilde f(s)$ turns to be the iterated Laplace transform 
of $K(r)$ we recognize that $\widetilde f(s)$ is the  Stieltjes transform of $K(r)$
and therefore the spectral distribution can be determined as the inverse Stieltjes transform of $\widetilde f(s)$ via the Titchmarsh inversion formula,
see {\it e.g.} \cite {Titchmarsh_1937,Widder_1946},
$$
\widetilde f(s) = \int_0^\infty \!\frac{K(r)}{s+r} \, \hbox{d} r\,,  
\quad
K(r) =
{\ds \mp \rec{\pi} \,
\hbox{Im} \left[\left. \widetilde f(s) \right|_{s=r e^{\pm i\pi}}\right]}\,. 
\eqno(2.11)$$
 As a consequence, in order to    prove  the  CM of the function in the L.H.S of Eq.(2.4) and determine   the corresponding spectral distribution,  its Laplace transform
 (expressed in two equivalent forms in the R.H.S. of Eq.(2.4))  will  be the starting point of our analysis.
\vsp
We recall that for the Mittag-Leffler functions in one and two-order parameter entering  Eqs.(2.3) and (2.2) respectively,  the conditions to be CM  on the negative real axis were  
derived  by Pollard \cite{Pollard 1948} in 1948, 
that is $0<\alpha\le  1$,
and  by Schneider \cite{Schneider 96} in 1996, that is $0<\alpha \le 1$ and 
$\beta\ge \alpha$. 
See also Miller and Samko \cite{Miller-Samko 97,Miller-Samko 01} for further details. 
\vsp
Before deriving the conditions of CM and the corresponding spectral function for  the Mittag-Leffler function in three parameters in Eq.(2.4),
 let us revisit the conditions of CM for the function in two parameters in Eq.(2.5)  following the approach by Gorenflo and Mainardi \cite{Gorenflo-Mainardi CISM97}.
 Since  the argument of our function  $at^\alpha$ must be negative we assume  $a=-1$
  (without loss of generality) so the corresponding Laplace transform  pair reads from Eq.(2.5),
  $$
t^{ \beta-1}\,{\ds E_{\alpha,\beta}(-t^\alpha)} \,\div \,
{\ds \frac{s^{-\beta}}{1+ s^{-\alpha}}}
={\ds \frac{s^{\alpha -\beta}}{  s^{\alpha}+1 }}
\,.
\eqno(2.12)$$
%% Excluding the trivial case $\alpha =\beta =1$ for which the function reduces to
%% the exponential $\exp(-t)$ 
We prove  the existence of the corresponding spectral distribution
using the complex Bromwich formula to invert the Laplace
transform.  
Taking $0\!<\!\alpha\! <\!1$, the denominator does not exhibit any zero in the main branch so, 
  bending the Bromwich path into the equivalent 
Hankel path (the well known  loop around the  negative real semi-axis for the reciprocal of the Gamma function),
we get
 $$  t^{\beta-1}\, E_{\alpha,\beta}(-t^\alpha)\!=\!  \int_0^\infty \!\!\e^{-rt}\, K_{\alpha, \beta} (r) \, \d r \,,\eqno(2.13)$$
 with
$$
%\begin{array}{ll}
K_{\alpha, \beta} (r)
% &
  = 
{\ds - \rec{\pi} \,
\hbox{Im} \left[\left. 
\frac{s^{\alpha-\beta}}{s^\alpha +1}\right|_{s=r e^{i\pi}}\right]}
% \\ \\
%&
 =
 {\ds \frac{r^{ \alpha -\beta} }{ \pi}\,
   \frac{\sin \,[(\beta -\alpha)\pi]  + r^\alpha \, \sin\,(\beta \pi)}
   { r^{2\alpha} + 2 r^{\alpha}\,\cos(\alpha\pi)+1}}\,.
%   \end{array}
   \eqno(2.14)$$
We easily recognize  
$$ K_{\alpha ,\beta}(r) \ge 0 \quad \hbox{if}\quad  0<\alpha\le \beta \le 1\,,
\eqno(2.15)$$
including  the limiting case $\alpha=\beta=1 $ where our Mittag-leffler function reduces to the exponential $\exp(-t)$ and $K_{1,1}(r) = \delta(r-1)$.
In fact, the denominator in Eq.(2.14) is non negative being greater or equal
to  $(r^\alpha-1)^2$ and the numerator is non negative as soon as the two sin  functions are both non-negative. 
 \vsp
 We note that the conditions (2.15)  on the parameters $\alpha$ and $\beta$ can also be justified by noting that in this case the resulting function is CM as a product of two CM functions. In fact $t^{\beta-1}$ is CM if $\beta<1$ whereas
 $E _{\alpha, \beta}(-t^\alpha)$ is CM if $0<\alpha\le 1$ and $\beta \ge \alpha$.  
 \vsp    
The particular case $\beta=1$ is recovered as
$$  E_\alpha(-t^\alpha)\!=\!  \int_0^\infty \!\!\e^{-rt}\, K_\alpha(r) \, \d r \,,\;
 K_\alpha(r) \!=\!  	\frac{r^{\alpha-1}}{\pi}\,
   \frac{  \sin (\alpha \pi)}
    {r^{2\alpha} + 2 r^\alpha \cos  (\alpha \pi) + 1}
	\,. \eqno(2.16)$$ 
Finally,   we devote to our attention 
to the more general function
$$ \xi_G(t) := t^{ \beta-1}\,{\ds E_{\alpha,\beta}^{\gamma}\left( -t^\alpha\right)}\,, \eqno(2.17)$$
with Laplace transform (as derived from (2.4) with $a=-1$)
$$
\widetilde \xi_G(s) = 
{\ds \frac{s^{-\beta}}{1+ s^{-\alpha}}}
=
{\ds \frac{s^{\alpha \gamma -\beta}}{\left( s^{\alpha} + 1\right)^\gamma}}\,, \eqno(2.18)
$$
where the notation $\xi_G(t)$  has been introduced for future convenience
in the applications to dielectrics.
\vsp 
In view of  Titchmarsh formula (2.11) applied to the 
equivalent Laplace transforms (2.18) we get
$$  
\xi_G(t) = \int_0^\infty \e^{-rt}\, K_{\alpha,\beta}^\gamma (r)\, {\mbox{d}}r \,,\eqno(2.19)
$$
with
$$
\begin{array}{ll}
K_{\alpha,\beta}^{\gamma} (r) 
&= 
{\ds \frac{r^{-\beta}}{\pi} \,{\mbox{Im}} \left\{ 
{\mbox{e}}^{i\beta\pi} \left( \frac{r^{\alpha} + {\mbox{e}}^{-i \alpha\pi}}
{r^{\alpha} + 2 \cos ( \alpha\pi)  + r^{-\alpha}}\right)^{\gamma} \right\}} \\ \\
&= 
{ \ds - \frac{r^{\alpha \gamma -\beta}}{\pi}\, {\mbox{Im}} \left\{
\frac{{\mbox{e}}^{i (\alpha\gamma-\beta)\pi}}
{(r^\alpha {\mbox{e}}^{i\alpha\pi}+1)^{\gamma}}
\right\}}\,.
\end{array}
\eqno(2.20)
$$
For $\gamma=1$ we obtain from Eq.(2.20) the spectral distribution 
 of the two parameter  Mittag-Leffler function outlined in (2.14).
For $\gamma\ne 1$ it may be cumbersome  to get an explicit expression for the spectral distribution so  we content ourselves with the two equivalent expressions in Eq.(2.20) 
in view of the fact that in any case  the distribution must be numerically computed. 
\vsp
By the way it is more relevant for us  to derive the conditions  on the parameters $\alpha,\beta, \gamma$ 
 required to ensure the non negativity of the spectral distribution  (2.20).  
In the following  we will show that the required conditions  consist in the inequalities
that we write in two equivalent forms, 
$$ 0<\alpha \le 1, \; 0< \beta \le 1, \; 0< \gamma \le \frac{\beta}{\alpha} 
\Longleftrightarrow 
0<\alpha \le 1,  \; 0< \alpha \gamma \le \beta \le 1, \eqno(2.21)$$ 
that for $\gamma =1$ reduce to the single inequality $0< \alpha \le \beta \le 1$ 
in Eq.(2.15) required for the two-parameter Mittag-Leffler function (2.12).
%%%%%%%%%%
\vsp
For this purpose
we take  advantage   of  the requirements  stated in the treatise
by Gripenberg et al. \cite{Gripenberg-et-al 90}, see  Theorem 2.6, pp. 144-145,
that provide necessary and sufficient conditions to ensure the CM of a function $f(t)$ 
based on its Laplace transform $\widetilde f(s)$.
%% \vsp
Hereafter, we recall this theorem by using our notation.
\vsp
{\bf Theorem} The Laplace transform  $\widetilde f(s)$ of a function $f(t)$ 
that is locally integrable on $\RR^+$ and CM  has the following properties:
\\
(i) $\widetilde f(s)$  an analytical extension to the region $\CC - \RR^-$;
\\
(ii) $\widetilde f(x) = {\widetilde f}^*(x)$ for $x\in (0,\infty)$;
 \\
 (iii) $ {\ds \lim_{x\to\infty} \widetilde f(x)} =0$;
 \\
 (iv) $\hbox{Im} \{\widetilde f(s)\} <0$ for $\hbox{Im} \{s\} >0$;
 \\
 (v) $\hbox{Im}   \{s\,\widetilde f(s)\} \ge 0$ for $\hbox{Im}\{s\}>0$ and
   $\widetilde f(x)\ge 0$ for $x \in (0,\infty)$.
   \\ Conversely, every function $\widetilde f(s)$ that satisfies (i)--(iii) together
   with (iv) or (v), is the Laplace transform of a function $f(t)$, which is locally integrable on $\RR^+$ 
   and CM on $(0,\infty)$. 
%% we can derive the conditions of CM
%% of the response functions  related in some way to the   Prabhakar generalized
%%Mittag-Leffler functions. For this purpose we must assume $a$ negative, in particular $a= -1$
%% and $\alpha, \beta, \gamma \le 1 $ in order to consider this property connected with the most common 
%% response functions in modelling dielectric relaxation, that is with the models referred to as
%%C-C, D-C, H-N and also Kohlrausch (stretched exponential).
%%%%%%%%%%%%%%%%%%%%%%%%%%%%%%%%%
\vsp
We recognize that  the requirements (i)--(iii) 
for $\widetilde \xi_G(s)$  are surely satisfied  with the first two conditions 
in the LHS of Eq.(2.21), that is $0<\alpha < 1$, $0<\beta < 1$   
but for any $\gamma >0$.  
So for us it suffices to determine which additional condition is implied from the requirement (iv).
We will prove that this relevant condition is just $0<\alpha \gamma - \beta \le 0$, namely 
$ 0< \gamma \le  \beta/\alpha$,  as stated in   Eq.(2.21). 
\vsp
Assuming the second expression of the Laplace transform in Eq.(2.18), the requirement (iv) reads:
$$
\Lambda(s) :=
{\mbox{Im}}\left[ \frac{s^{\alpha \gamma - \beta}}{(s^{\alpha}+1)^{\gamma}} \right] < 0 \quad {\rm where} \quad {\mbox{Im}}\{s\} > 0\,.\eqno(2.22)
$$
Setting  $s=r e^{i\phi}$  in the complex upper half-plane (Im$\{s\}>0$)    
we  consider
$$
\Lambda (r,\phi) :=
{\mbox{Im}}\left[
\frac{(r \e^{i\phi})^{\alpha \gamma - \beta} 
(1+r^{\alpha} \e^{-i\alpha\phi })^{\gamma}}
{\left|1 + r^\alpha  \e^{i\alpha\phi }\right|^{2\gamma}}
 \right] \; \mbox{with}\;  r>0 , \; 0 < \phi < \pi.
 \eqno(2.23)
$$
To prove that  $\Lambda (r,\phi)$ is negative it is sufficient to consider
the numerator because the denominator is always non-negative. 
Setting
$$ z = (r \e^{i\phi})^{\alpha \gamma - \beta} 
\, (1+r^{\alpha} \e^{-i\alpha\phi })^{\gamma}
= \rho \, \e^{i\Psi}\,,
\eqno (2.24)$$
we must verify that the conditions on $\{\alpha, \beta, \gamma\}$  stated in Eq.(2.21)
ensure that $z$ has negative imaginary part so it is located in the lower half plane with 
$$ -\pi <\Psi <0\,.\eqno(2.25)$$
Let   
$$
z_1 = r^{\alpha \gamma - \beta} \,\e^{i(\alpha \gamma - \beta)\phi} =
\rho_1\,\e^{i\Psi_1}\,,
\quad 
\rho_1 = r^{\alpha \gamma - \beta},\; 
 \Psi_1 = (\alpha \gamma - \beta)\phi \,,
\eqno(2.26)
$$
$$
z_2 = r^{\alpha}\, \e^{-i\alpha\phi} = \rho_2 \,  \e^{i\Psi_2}\,,
\quad  
\rho_2 = r^{\alpha}, \;
 \Psi_2 = -\alpha\phi \,,
 \eqno (2.27)$$
and 
$$
z_3 = (1+z_2)^{\gamma} = \rho_3 \,\e^{i\Psi_3}\,,
\quad  
\rho_3 =|1+ r^{\alpha}\, \e^{-i\alpha\phi}|^\gamma, \;
  -\alpha\gamma \phi <\Psi_3<0\,,
\eqno(2.28)$$
so  we can write the complex number in  Eq.(2.24) as 
$$
z= z_1 \cdot z_3 = \rho_1 \,\e^{i\Psi_1} \rho_3 \, \e^{i\Psi_3} = \rho\, \e^{i\Psi}
\quad \mbox{with} \quad
\rho = \rho_1 \rho_3, \; \Psi = \Psi_1 + \Psi_3\,. 
\eqno(2.29)
$$
Now assuming  $0<\phi<\pi$  a we find
for $\alpha \gamma - \beta <0$:
$$ -(\beta -\alpha \gamma) \pi <\Psi_1<0\,, \eqno (2.30)$$
$$ -\alpha \gamma \pi < \Psi_3<0\,. \eqno(2.31)$$
For $\alpha \gamma - \beta =0$ we find 
$ \Psi_1 =0$ and $ -\alpha \gamma  \pi= -\beta \pi  < \Psi_3<0$.
As a consequence for $\alpha \gamma - \beta \le 0$, 
 by summing ($\Psi =  \Psi_1 + \Psi_3$), we finally get 
$$ -\pi < -\beta \pi < \Psi <0\,,\eqno(2.32) $$
so the inequality (2.25) is proved since  $0<\beta<1$.

\newpage
%%%%%%%%%%%%
\section{Mathematical models for dielectric \\relaxation}
We intend to discuss  how  the Laplace transform pair outlined in Eq. (2.18) coupled with the conditions
(2.21) on the 3-order parameters   can  be considered as  the pair $\xi_G(t)$ 
($t\ge 0$) and $\widetilde \xi_G(s)$
($s=i\omega$) for a possible mathematical model of  the response function and the complex susceptibility
in the framework of a general relaxation theory   of dielectrics.
\vsp
We first show how the three classical models referred to Cole--Cole (C-C), Davidson--Cole (D-C)
and Havriliak--Negami (H-N)  are contained in our general model described by a response function expressed in terms the three-parameter Mittag-Leffler function, see Eqs.(2.17), (2.18) subjected to the conditions (2.21)
with $ \alpha \gamma - \beta=0$, 
according to the scheme      
$$ 
%% \alpha \gamma - \beta =0 \; \hbox{with} \; 
\left\{
\begin{array}{lllll}
{\ds 0<\alpha<1\,, \, \beta =\alpha\,,\, \gamma=1} &&& {\mbox{C-C}} \; \{\alpha\}\,,\\
{\ds \alpha=1\,, \, \beta = \gamma \,,\, 0<\gamma<1} &&& {\mbox{D-C}} \; \{\gamma\}\,,
\\
{\ds 0<\alpha<1\,, \,\, 0< \gamma<1} &&& {\mbox{H-N}} \; \{\alpha,\gamma\}\,.
\\
\end{array} 
\right.
 \eqno (3.1) 
$$
%\vsp
Then, we consider some study-cases when the inequality $0< \alpha \gamma \le  \beta$
holds provided  $0<\alpha \le 1$, $ 0<\beta \le 1$ in agreement of Eq.(2.21).
However, we have restricted our attention to values of  $\gamma \in (0,1]$. 
  For this purpose we  exhibit  3D plots for the response function $\xi_G(t) $
 %% , see Eq.(2.17), 
and the corresponding spectral distribution $K_{\alpha, \beta}^{\gamma}(r)$,
 %5 see Eq.(2.20), 
keeping fixed two of the three-order parameters.   
 
\subsection{The classical dielectric functions}
 \paragraph{The Cole--Cole relaxation model.}
The C-C relaxation model is a non-Debye relaxation model depending  on one parameter, 
say  $\alpha \in (0,1)$, see \cite{Cole-Cole 41,Cole-Cole 42},
 that for $\alpha=1$ reduces to the standard Debye model.
 The corresponding complex susceptibility ($s=-i\omega$) and response function read
$$
\widetilde \xi_{{\mbox{C-C}}} (s) \!=\! \frac {1}{1 + s^\alpha } \div\ 
\xi_{{\mbox{C-C}}}(t) \!=\! t^{\alpha-1} E_{\alpha, \alpha}^1 (-t^\alpha)
\! =\!  
- \frac{\d}{\d t} E_\alpha(-t^\alpha), \; 0<\alpha < 1. \eqno(3.2)
$$
 This case is obtained for $\{0<\alpha=\beta <1,\; \gamma=1\}$ so that
  $\alpha \gamma - \beta =0$. 
%%%%%%%%5
\paragraph{The Davidson--Cole relaxation model.}
The D-C relaxation model is a non-Debye relaxation model depending  on one parameter, say $\gamma\in (0,1)$,
see \cite{Davidson-Cole 51}, that for $\gamma=1$ reduces to the standard Debye model.
The corresponding complex susceptibility  and response function read
$$ 
\widetilde \xi_{{\mbox{D-C}}} (s) 
\!= \!\frac {1}{\left(1 + s\right)^\gamma } \div 
\xi_{{\mbox{D-C}}}(t)
\!=\!  t^{\gamma-1} E_{1, \gamma}^\gamma (-t)
\! =\!  
\frac{t^{\gamma-1}}{\Gamma(\gamma)}\\e^{-t} , \; 0<\gamma < 1. \eqno(3.3)
$$
This case is obtained for $\{\alpha=1, \; 0<\beta=\gamma<1\}$ so that $\alpha \gamma - \beta=0 $ again. 
 \paragraph{The Havriliak--Negami relaxation model.}
The H-N relaxation model is a non-Debye relaxation model depending  on two parameters, 
$\alpha \in (0,1)$  and $\gamma \in (0,1)$,
see 
\cite{Havriliak-Havriliak POLYMER96,%
Havriliak-Negami JPSC66,%
Havriliak-Negami POLYMER67}, 
that
for $\alpha =\gamma =1$ reduces to the standard Debye model.
The corresponding complex susceptibility  and response function read 
$$ 
\widetilde \xi_{{\mbox{H-N}}} (s) = \frac {1}{\left(1 + s^\alpha \right)^\gamma }
\,\div\,\xi_{{\mbox{H-N}}} (t) = t^{\alpha \gamma-1} \, E_{\alpha,\alpha \gamma}^{\gamma}(-t^\alpha)\,,
\;  0<\alpha,\gamma < 1\,. \eqno(3.4)
$$
This case is obtained for $\{0<\alpha <1, \;0<\gamma <1, \; \beta = \alpha \gamma\}$ so that   $\alpha \gamma -\beta=0$ again.
 %\vsp 
We note that  this model
%%the H-N relaxation model 
for $\alpha \in (0,1)$ and  $\gamma=  1$ reduces
to the C-C model, while for $\alpha  = 1$ and $\gamma \in (0,1)$ to the D-C  model.
% \vsp
We also recognize that whereas for the C-C and H-N models the corresponding 
response functions decay like a certain negative power of time
(namely $t^{-\alpha -1}$ for a Tauberian theorem), 
the D-C response function exhibits an exponential decay being $\alpha=1$.
%% The latter is often used to represent the asymmetric  peak in glass-forming materials, in 
%% contrast to the C-C relaxation modeling symmetric non-Debye peaks of 
%% the imaginary part of the complex dielectric permittivity.   
%% \\ Acording to Hanyga and Seredy{\'n}ska \cite{Hanyga JSP08},
%% the relaxation models of C-C, D-C and H-N are distinguishable by inspection of their Cole-Cole plots. 
%%%%We also note that D-C relaxation differs from the C-C and H-N by an exponential  decay. 
%%%%% In fact, for each of these relaxation models the relaxation function can 
%% be expressed in terms of a complex contour integral along a Hankel loop $Ha$ 
%% encircling the cut on the negative real axis and hence to the Laplace transform 
%% of the jump of the complex dielectric permittivity on the cut.
%%%%%%%%
\subsection{Survey of the general response functions with their spectral distributions}
%%%%%%%%%%%
In order to visualize the effects of varying the three order-parameters in our general model 
we survey  some  particular cases
by exhibiting separately the response functions $\xi_G(t)$ for $0<t<10$ 
given by Eq.(2.17) 
and the corresponding spectral distributions for $0<r<10$ given by Eq.(2.20).
%% in 3D plots.
%% As previously outlined, 
For this purpose we provide 3D plots by  keeping fixed two of the three positive 
order-parameters $\{\alpha,\beta,\gamma\}$, all of them less than unity and 
subjected to the condition $\alpha \, \gamma <\beta$.
%\vsp     
In Figs.1 and 2,  for fixed $\alpha =1/2$ we compare versus $\gamma \in (0,1)$ 
the plots of $\xi_G(t)$  
and  of the corresponding spectral distribution for $\beta=\gamma/2, 2\gamma/3$.    
%% \vsp
In Figs.3 and 4,  for fixed $\gamma=1/2$ we compare versus $\alpha \in (0,1)$ the plots of $\xi_G(t)$ 
and  of the corresponding spectral distribution
for $\beta=\alpha/2, 2\alpha/3$.
% \vsp 
In Figs.5 and 6, for fixed $\alpha=1/2, \gamma=1$ and $\alpha=2/3, \gamma=1/2$
we compare versus $\beta \in (1/3,1)$ the plots of $\xi_G(t)$
and  of the corresponding spectral distribution. 
%% \vsp
%% Finally, in Figs. 4, 5, 6 we plot the spectral distributions corresponding to the responses functions 
%% in Figs. 1, 2, 3, respectively. 
      % Give a unique label
\newpage
\bigskip
\begin{figure}[h!]
\resizebox{0.80\columnwidth}{!}{\includegraphics{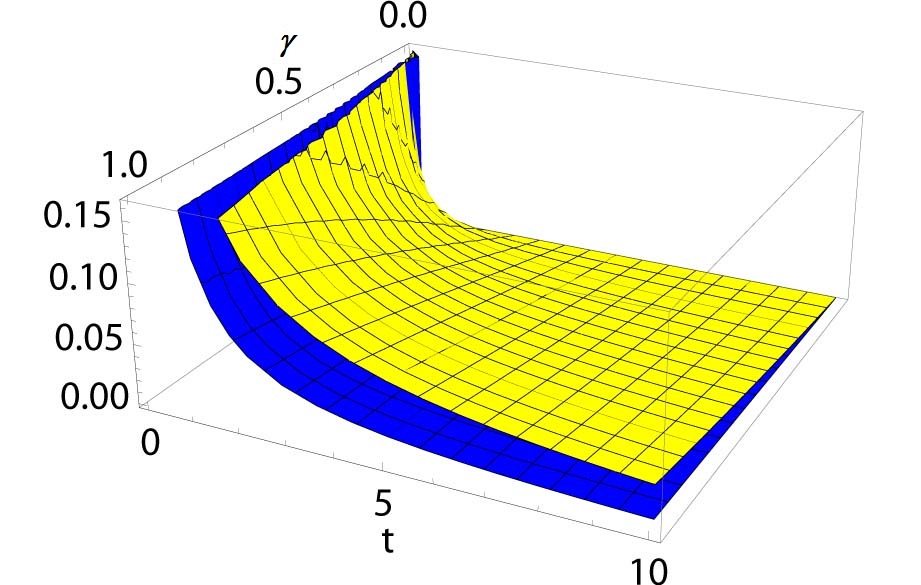} }
\caption{Plot of the response function $ \xi_G(t)$ 
versus $\gamma\in (0,1)$ and $t\in (0,10)$
for $\{\alpha=1/2, \beta=\gamma/2,\gamma\}$  (dark grey/blue surface) 
and  $\{\alpha=1/2, \beta=2\gamma/3,\gamma\}$ (light grey/yellow surface).
}
\label{fig:1}     
\end{figure}
  % Give a unique label
\bigskip
\begin{figure}[h!]
% \resizebox{0.75\columnwidth}{!}{\includegraphics{K_varying_gamma.eps} }
\resizebox{0.80\columnwidth}{!}{\includegraphics{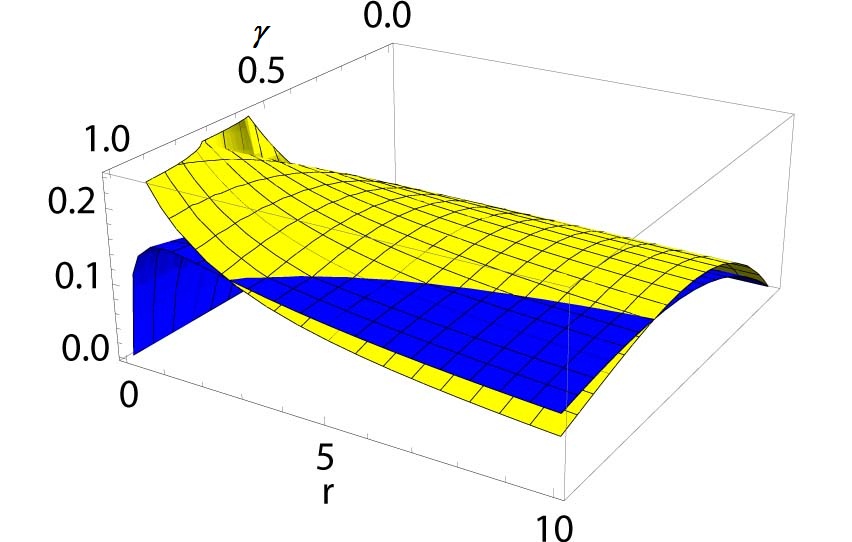} }
\caption{Plot of the spectral distribution  
$K_{1/2,\gamma/2}^\gamma(r)$ (dark grey/blue surface) and 
$K_{1/2,2\gamma/3}^\gamma(r)$ (light grey/yellow surface) 
for $0 < \gamma < 1$  and $0\leq r \leq 10$.}
\label{fig:2}       % Give a unique label
\end{figure}

\newpage

\bigskip

\begin{figure}[h!]
%%%%%%%%%%%
\resizebox{0.80\columnwidth}{!}{\includegraphics{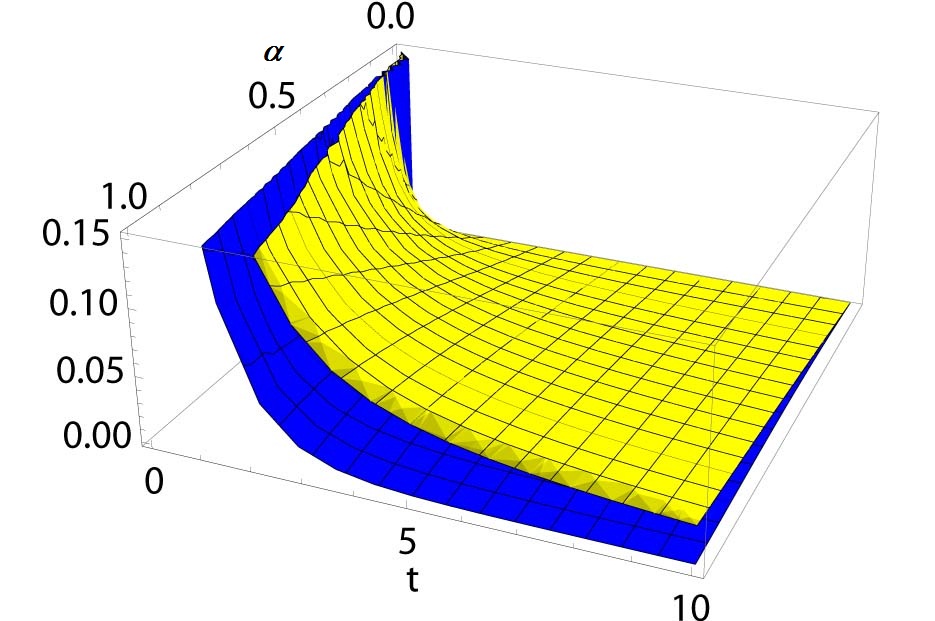} }
\caption{Plot of the response function $ \xi_G(t)$ versus $\alpha\in (0,1)$ and $t\in (0,10)$
for $\{\alpha, \beta=\alpha/2,\gamma=1/2\}$ (dark grey/blue surface) 
and   $\{\alpha, \beta=2\alpha/3,\gamma=1/2\}$ (light grey/yellow surface).
}
\label{fig:3} 
\end{figure}

\bigskip

%%%%%%%%%%
\begin{figure}[h!]
% \resizebox{0.75\columnwidth}{!}{\includegraphics{K_varying_alpha.eps} }
\resizebox{0.80\columnwidth}{!}{\includegraphics{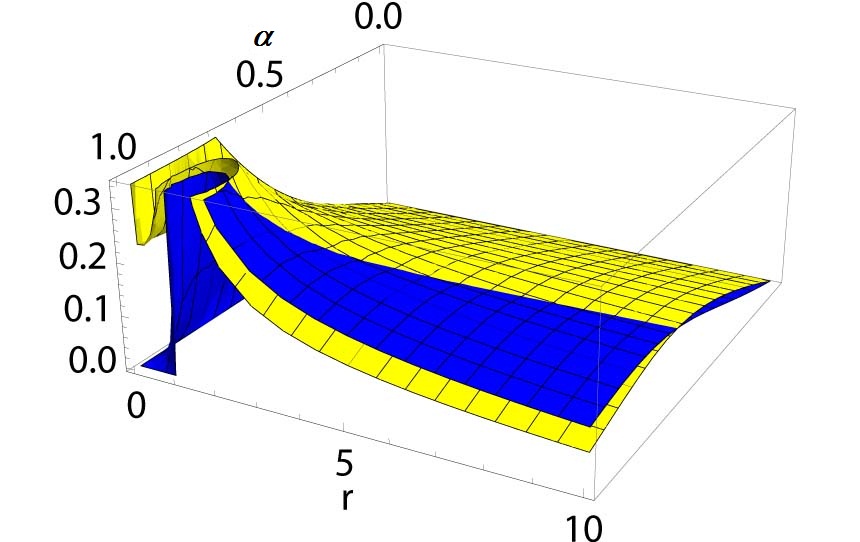} }
\caption{Plot of the spectral distribution 
$K_{\alpha,\alpha/2}^{1/2}(r)$ (dark grey/blue surface) and 
$K_{\alpha,2\alpha/3}^{1/2}(r)$ (light grey/yellow surface) 
for $0 < \alpha < 1$  and $0\leq r \leq 10$.}
\label{fig:4}       % Give a unique label
\end{figure}
%%%%%%%%%%%%%%%%%%
\newpage

\bigskip

\begin{figure}[h!]
% \resizebox{0.75\columnwidth}{!}{\includegraphics{function_varying_beta.eps} }
\resizebox{0.80\columnwidth}{!}{\includegraphics{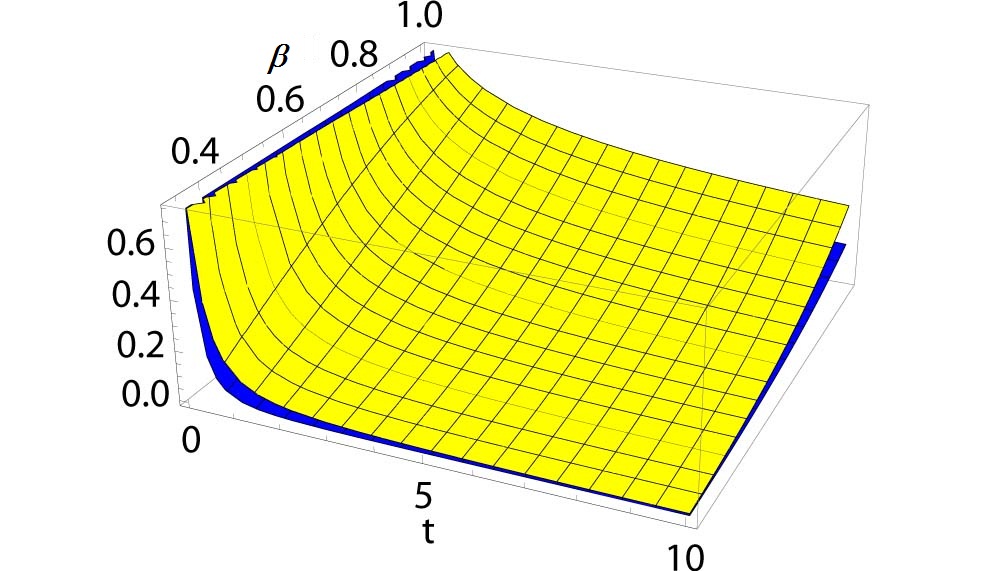} }
\caption{Plot of the response function $ \xi_G(t)$ versus $\beta\in (1/3,1)$ and $t\in (0,10)$
 for $\{\alpha=1/2, \beta,\gamma=1\}$ (dark grey/blue surface) 
and   $\{\alpha=2/3, \beta=2\alpha/3,\gamma=1/2\}$ (light grey/yellow surface).
}
\label{fig:5}       % Give a unique label
\end{figure}
%%%%%%%%%%%

\bigskip

%%%%%%%%%%%%%%%%%
\begin{figure}[h!]
% \resizebox{0.75\columnwidth}{!}{\includegraphics{K_varying_beta.eps} }
\resizebox{0.80\columnwidth}{!}{\includegraphics{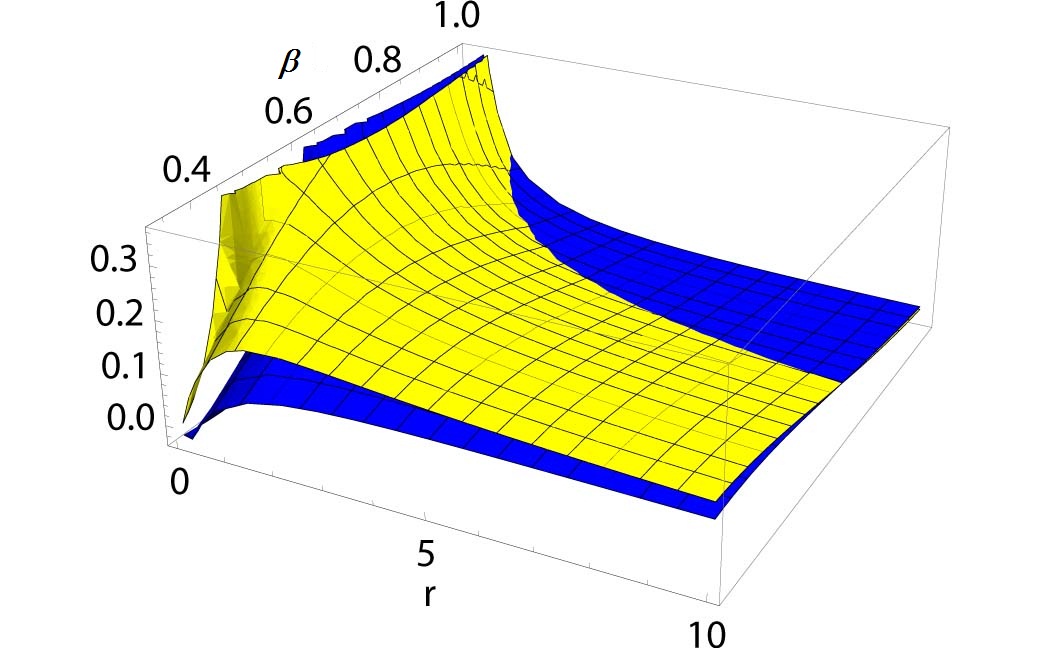} }
\caption{Plot of the spectral distribution 
$K_{1/2,\beta}^{1}(r)$ (dark grey/blue  surface) and 
$K_{2/3,\beta}^{1/2}(r)$ (light grey/yellow surface) 
for $1/3 < \beta < 1$  and $0\leq r \leq 10$. 
Note that $K_{2/3,\beta}^{1/2}(r) < 0$ when the condition $\beta \geq \alpha\gamma$ is not satisfied.}
\label{fig:6}       % Give a unique label
\end{figure}

\bigskip

\newpage
\vsp
For our plots MATHEMATICA (Wolfram) was used.  As far as we know, up to now  no official routine exists for the Prabhakar function whereas for the two parameter  Mittag-Leffler function  a number of authors have provided specific routines and/or numerical methods, see e.g.
\cite{Garrappa-Popolizio_ACM2013,GoLoLu_FCAA2002,%%
Hilfer-Seybold_ITSF2006, Seybold-Hilfer_FCAA2005,%%
Seybold-Hilfer_SIAM2008,Verotta_JPP2010,Zeng-Chen_ARXIV2013}.   

\section{Conclusions}
In this revised paper we have presented a quite general mathematical model 
for dielectrics that exhibit deviations from the standard relaxation Debye law.
This model is  based on a time response function expressed in terms of a Mittag-Leffler function 
with three order-parameters.  A restriction on these parameters is required 
to ensure  its  complete monotonicity  for $t>0$, so that the  resulting  relaxation process          
can be seen as a continuous distribution of elementary exponential processes by means of a
 corresponding  (non-negative) spectral distribution function. 
 \vsp
Our approach  allows one to derive from a unique mathematical framework
the  classical Cole-Cole, Davidson--Cole and Havriliak--Negami models 
that are usually adopted in the literature. However,  other laws can be 
 derived from our model that could better fit  some experimental data.
 \vsp      
 For some study--cases we have exhibited  plots  of the response functions (versus $t$) along 
  with their  corresponding (non-negative) spectral distributions (versus $r$) keeping fixed two of the three 
order-parameters $\{\alpha, \beta,\gamma\}$ of our Mittag-Leffler function.
We have verified  that when the three conditions stated in Eq.(2.21):
$0<\alpha \le 1$, $0<\beta \le 1$  and  $0<\gamma \le \beta/\alpha$
 are not all satisfied
the corresponding spectral distributions $K_{\alpha, \beta}^\gamma(r)$  turn out to be negative in some ranges of $r$. 
\vsp
As a matter of fact with this revised version   we have shown a noteworthy result   that improves   the conditions outlined   by Mainardi in his 2010 book \cite{Mainardi BOOK10}  and then  by us in our  original  paper published in 
 {\it Eur. Phys. J.-Special Topics}, Vol. 193, pp. 161--171 (2011). 
With respect to the published versions, we note that condition 
$0<\gamma \le 1$ can be relaxed provided that all  inequalities in Eq.(2.21)
hold true. 
  %% \cite{CMV_2011}.
\vsp
We point out that we were inspired  by the approach 
by Hanyga and Seredy{\'n}ska \cite{Hanyga JSP08} who 
have applied the theorem by Gripemberg et al \cite{Gripenberg-et-al 90}
in order to prove the CM of the 3-parameter Prabhakar function in the particular case $\beta=1$. 
 %\vsp
% In a future paper we will be  interested to  provide the fractional evolution equation % governing our general response function extending the results obtained  by
% Nigmatullin and Ryabov \cite{Nigmatullin-Ryabov 1997},
% Novikov et al. \cite{Novikov MS05} 
% and by Sibatov et al. \cite{Sibatov-et-al 2012} for the Havriliak--Negami model. 

\newpage
\section*{Acknowledgements}
This work  been carried when FM was a Visiting Professor  at the Department of Applied Mathematics,
IMECC, University of Campinas (Brazil), as a recipient of a Fellowship of the FAEPEX 184/10. %Government of Brazil.   
FM  appreciated the scientific atmosphere and perfect conditions for providing
research facilities at this Department.
\vsp
As far  the revised versions of this paper are concerned,  the authors are grateful to  Professors  Karina Weron, Rudolf Gorenflo, Tibor K. Pogany   and \v{Z}ivorad Tomovski for useful criticism.

\end{document}